\documentclass[man,12pt,floatsintext]{apa7}

\usepackage{apacite}
\usepackage{enumitem}
\usepackage{multirow}
\usepackage{placeins}
\usepackage{makecell}
\usepackage{subcaption}
\usepackage{float}

\title{An Evaluation Study of 2D and 3D Teleconferencing for Remote Physical Therapy}
\shorttitle{Evaluation of Spatial Remote Communication Systems}

\authorsnames[1,2,1,2,2,1]{Hanseul Jun, Husam Shaik, Cyan DeVeaux, Michael Lewek, Henry Fuchs, Jeremy Bailenson}
\authorsaffiliations{{Stanford University}, {University of North Carolina at Chapel Hill}}

\abstract{

The present research investigates the effectiveness of using a telepresence system compared to a video conferencing system and the effectiveness of using two cameras compared to one camera for remote physical therapy. We used Telegie as our telepresence system, which allowed users to see an environment captured with RGBD cameras in 3D through a VR headset. Since both telepresence and the inclusion of a second camera provide users with additional spatial information, we examined this affordance within the relevant context of remote physical therapy.  Our dyadic study across different time zones paired 11 physical therapists with 76 participants who took on the role of patients for a remote session. Our quantitative questionnaire data and qualitative interviews with therapists revealed several important findings. First, after controlling for individual differences between participants, using two cameras had a marginally significant positive effect on physical therapy assessment scores from therapists. Second, the spatial ability of patients was a strong predictor of therapist assessment. And third, the video clarity of remote communication systems mattered. Based on our findings, we offer several suggestions and insights towards the future use of telepresence systems for remote communication. }

\begin{document}
\maketitle

\section{Introduction}
Within the landscape of remote communication, video conferencing (VC) systems have grown in popularity. Through sharing both visual and auditory information, VC systems can enhance the experience of traditional phone calls ~\cite{kraut1995prospects}. The capabilities of these systems have been shown to be useful in multiple interpersonal contexts. However, one area that VC lacks in is providing spatial information. Users are only able to view one angle of their communication partner based on the position of their camera.

For this reason, three-dimensional telepresence (TP) systems may represent the next iteration of commonly used remote communication technology. Telepresence is characterized by a sensation of presence in an environment that is mediated by a system ~\cite{steuer1992defining}. Whereas VC systems exhibit spatial limitations, TP systems allow users to move around their digitally-mediated communication environment. Therefore, users in separate physical locations are able to view each other from multiple viewpoints. 

The present study investigates TP as a potential successor to VC within the context of physical therapy. Spatial information is important to physical therapists, who often have to move around patients during face-to-face sessions. Therefore, the affordances of TP may prove to be useful for sessions that take place remotely. We examine this through an empirical study that compares the effectiveness of TP and VC during remote physical therapy sessions. We recruited 11 therapists and during each session, the therapists provided instructions of six exercises. The exercises (e.g., squat~\cite{suciu2016gait}) included spatial aspects. We also test how the inclusion of a second camera to both TP and VC might enhance the outcomes of these sessions by providing additional spatial information. Telegie\footnote{\url{https://telegie.com}} was used as the TP system for the remote communication between therapists and patients. The system supports real-time communication utilizing AR/VR headsets and RGBD cameras~\cite{jun2018easy}. The system supports a wide range of devices utilizing web technology.

\section{Related Work}
\subsection{Examples of Telepresence Systems}
In this section, we describe previous examples of TP systems that use RGBD cameras and AR/VR headsets.

In 2011, \citeA{maimone2011encumbrance} implemented a TP system using RGBD cameras and autostereoscopic monitors. In a subsequent iteration of the system, the autostereoscopic displays were replaced with AR headsets~\cite{maimone2013general}. Projectors were used to supplement the brightness of AR headsets as displays.

In 2012, \citeA{steed2012beaming} introduced Beaming, a TP system that supports an asymmetric setting of a single person beaming into a group of people. This asymmetric TP system has a VR system, called the transporter, that allows one user to face many people on the other side.

In 2013, \citeA{beck2013immersive} presented a group-to-group TP system with projectors and depth cameras. Their system supports connection between groups across two different locations by reconstructing the groups into meshes. By wearing shutter glasses, users can stereoscopically see the reconstructed meshes of the other group.

In 2015, \citeA{roberts2015withyou} introduced the \textit{withyou} system. Users of their system stand inside cubic, projector-based immersive displays where they are surrounded by multiple cameras. Within the cubes, the users can see the reconstructed version of others for remote communication. Their system supports stereoscopic rendering with the use of stereo glasses and more than two cubic displays.

Also in 2015, \citeA{kowalski2015livescan3d} introduced LiveScan3D, an open-source 3D data acquisition system using multiple Kinect 2 devices. Their goal was to provide an open-source reconstruction system based on multiple cameras that is easy to use. In 2017, the same scholars released an application for HoloLens devices that received and rendered point clouds from LiveScan3D\footnote{\url{https://github.com/MarekKowalski/LiveScan3D-Hololens}}. The use of an AR headset as the viewer for reconstructed scenes makes this a TP system. Source code for LiveScan3D and its extension are both available.

In 2016, Room2Room~\cite{pejsa2016room2room} and Holoportation~\cite{orts2016holoportation} were introduced. Room2Room allows remote communication between people in two different rooms. It captures a room using an RGBD camera and displays it in the other room using a projector. In their study where participants were asked to perform a collaborative assembly task, the researchers found their system superior to video chat in completion time, presence, and communication efficiency. Holoportation captures users with multiple RGBD cameras and reconstructs them into high-quality textured meshes. By displaying these meshes on AR headsets, the system facilitates dyadic remote communication.

In 2018, \citeA{kolkmeier2018little} introduced OpenIMPRESS, a software toolkit for mixed reality remote collaboration systems. With this system, a person can wear a VR headset with a hand tracking system attached to communicate with a person at a remote location who is wearing an AR headset.

In this paper, we use a TP system that supports both VR and AR headsets through web technology and requires only one RGBD camera for its operation. Multiple RGBD cameras may be used to increase the surface area that the TP system can capture and show to other users.

\subsection{Prior Research on  Virtual Reality for Physical Therapy}

Although there are many VR applications built for physical therapy, not many of them use VR for TP. Most of them consist of solo VR experiences where physical therapy patients receive instructions to complete on their own. A subset of these systems allows remote therapists to monitor physical therapy patients. Many of the applications introduced in this section use hardware apart from headsets for VR (e.g, rendered scenes displayed on a desktop computer). In these cases, the researchers used VR to display 3D content, but did not leverage the higher degree of immersion afforded by head-mounted displays.

\citeA{golomb2009eleven} conducted a 6 to 11-month clinical pilot study on using VR telerehabilitation for the treatment of three adolescents with hemiplegic cerebral palsy. To use this system, patients wore sensing gloves that could track their hand movements. These gloves allowed patients to practice moving their hands while following gamified instructions from computer monitors. Qualitative results from the pilot study demonstrated that "remote electronic monitoring is not enough; humans must be heavily
involved in remote monitoring. Human contact and human understanding are key to the success of telerehabilitation"~\cite[p.~27]{golomb2009eleven}.

VR has not only been used for providing therapy instructions, but also for building therapy programs. \citeA{camporesi2013vr} built a VR system that allows therapists to create new therapy programs intuitively. Using the Kinect, their system can capture movements from the therapist, record them as a therapy program, and automatically deliver this program to patients. Therapists can also use this system to monitor the activity of the patients. This work has been improved in subsequent research~\cite{kallmann2015vr} to allow patients to see the difference between their own movements and therapists' recorded movements by visualizing joint angle errors.

In their effort to apply VR to treating Parkinson's Disease patients, \citeA{feng2019virtual} compared VR-based therapy to conventional physical therapy in their 12-week study with 28 Parkinson's Disease patients. At the end of the 12-week period, researchers found that the group that used VR outperformed the group that received conventional physical therapy in terms of balance and gait.

While the existing systems are primarily for solo use, our study will be based on a system built for remote communication between therapists and patients. Our system also leverages the immersiveness of VR headsets.

\section{Telegie System Overview}
The Telegie system consists of two applications: the transmitter and the viewer. The transmitter sends RGBD streams from the cameras to the viewers, and the viewer renders the incoming RGBD streams to the users. The transmitter is supported by Windows 10 computers connected to Azure Kinect and iPhone devices. The viewer was built as a web application that can run on any device with a web browser, including VR headsets. The majority of the codebase is written in C++ and is shared between the transmitter and the viewer. Emscripten\footnote{\url{https://emscripten.org/}} is used to compile the functionalities into WebAssembly for their use in the viewer.

A TP call between two people---equivalent to a video conference call---happens in three steps. First, one user creates a room using their transmitter and obtains a room ID. Next, the other user joins the same room with their transmitter using the obtained room ID. Finally, both users enter the room using their viewers and start seeing each other.

Broadcasting can happen in two steps. The broadcasting user can create a room using a transmitter and obtain a room ID, then other users can enter the room through the website that shows the list of rooms.

It is also possible for a user to leverage multiple transmitters to increase the visual quality of a TP call. Additional transmitters can be added by joining the same room created by the first transmitter. After calibration, the additional transmitter can provide more information to the viewers. This manual calibration process is based on the additional transmitter's relative position and rotation to the first transmitter. This feature has been added for the present study.

Users of the system were able to view their remote interlocutor from the different angles captured by the cameras placed next to them. Viewing these different angles required physically walking around the TP system’s virtual environment.

\subsection{Number of External Cameras}
In order for a TP system to work, there must be a way for the system to capture its users to render them in front of others~\cite{maimone2011encumbrance}. While operating without any external cameras would be ideal, the cameras attached to AR and VR headsets, which are affixed to a user’s head, are too poorly positioned to capture sufficient visual information about a user's own body. For this reason, we chose to have an external RGBD camera. We required only one external RGBD camera per user to keep the installation of the system as simple as possible. As demonstrated in the present study, using multiple cameras is supported to provide better visual quality~\cite{orts2016holoportation}, but only as an option, not as a requirement.

\subsection{Distances between People}
TP systems can and should set the distance between people. Personal space literature~\cite{hayduk1983personal} sheds light on how social context is a major factor in deciding what distance is the most preferable. For example, people prefer to maintain larger distances towards strangers than towards their friends and significant others. This is relevant to TP, as AR studies have shown that people behave similarly towards virtual humans as they do towards real people~\cite{lee2018effects, miller2019social}. Based on the contextual nature of personal space preferences, we decided to provide users with an interface to set the distance.

\subsubsection{Matching Floors}

\begin{figure}[H]
  \centering
  \includegraphics[width=0.8\textwidth]{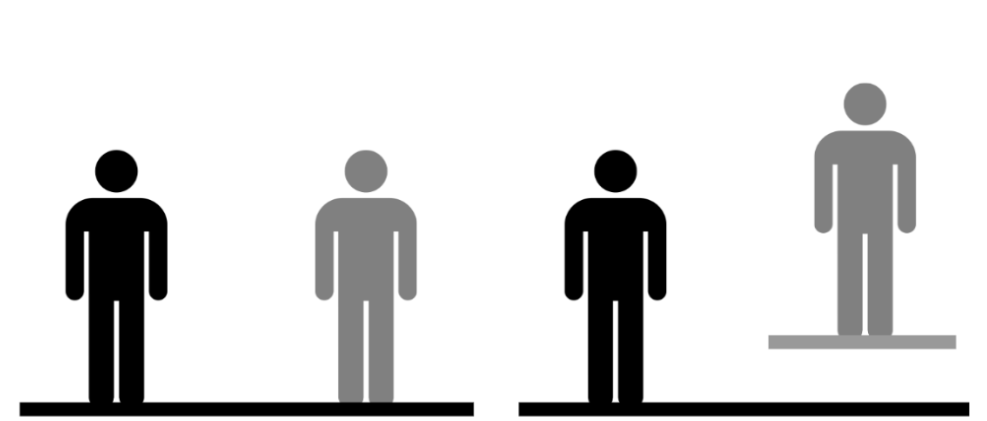}
  \caption{People on the same floor level (left) and people on different floor levels (right).}
  \label{fig:floors}
\end{figure}

When people meet each other in the real world, they are usually standing on the same floor. Therefore, we decided to replicate this experience in our system. Our system does not allow rising, sinking, or tilting floors. Matching floors involved setting constant y values and rotation values except for yaw. Figure~\ref{fig:floors} depicts matching floors and non-matching floors.

\subsection{Transmitter-Viewer Pipeline}

\begin{figure}[H]
  \includegraphics[width=\textwidth]{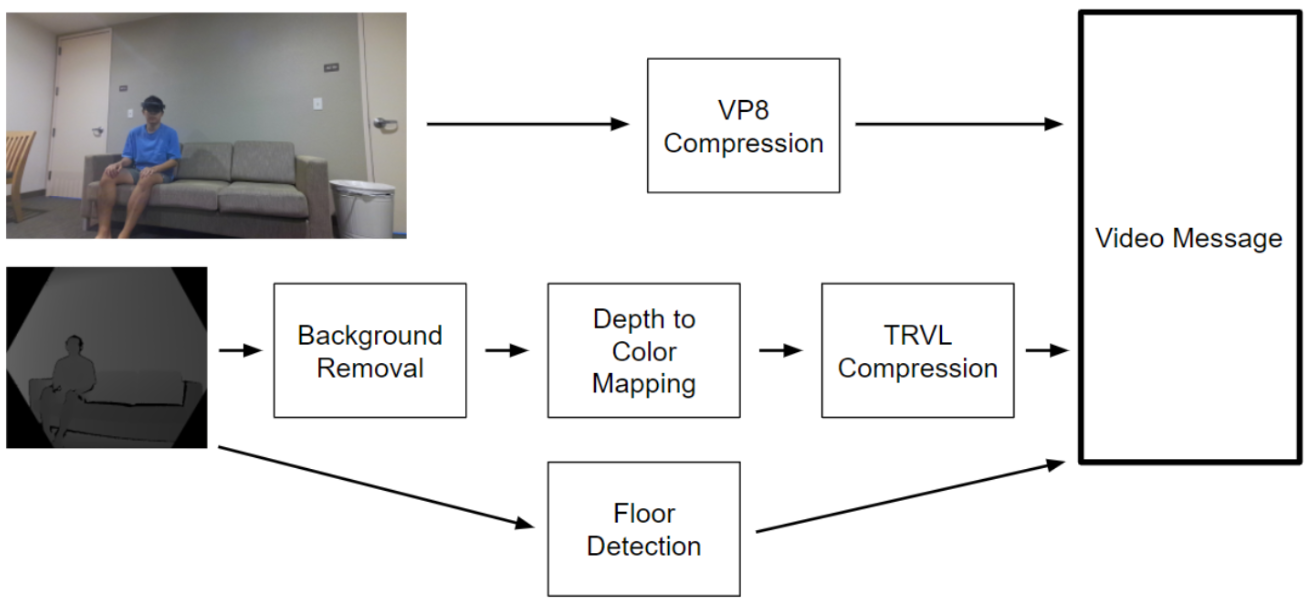}
  \caption{Construction of a video message from the color and depth pixels of an RGBD camera frame.}
  \label{fig:pipeline}
\end{figure}

The transmitter-viewer pipeline of Telegie delivers color, depth, floor, and audio information from a transmitter to a viewer. Color, depth, and floor information form a video message for every camera frame. For color information, color pixels are encoded in VP8 using libvpx\footnote{\url{https://chromium.googlesource.com/webm/libvpx}}. For depth information, depth pixels go through optional background removal, map to the color camera's coordinate system, and undergo Temporal RVL compression~\cite{jun2020temporal}. Floor information gets extracted from the depth pixels. Audio information is encoded in the Opus codec\footnote{\url{https://opus-codec.org/}} and gets sent separately from video messages. Figure~\ref{fig:pipeline} describes the construction of video messages.

\subsection{Networking Between Transmitter and Viewer}
For users to see each other, network packets including video messages are sent from transmitters to viewers. To support viewers running on web browsers and connections outside of local networks and across routers and firewalls, we utilized libdatachannel\footnote{\url{https://github.com/paullouisageneau/libdatachannel}}---an implementation of WebRTC data channels. Unreliable data channels were used for real-time communication with lower latency. Unfortunately, the use of these channels introduced packet loss. 

To handle packet loss, our system adopted a fountain code---Wirehair\footnote{\url{https://github.com/catid/wirehair}}. From a set of packets, a fountain code can provide limitless packets for packet loss recovery. From the receiving side, the original set of packets can be recovered after receiving sufficient fountain code packets. After splitting video messages into packets and encoding them in Wirehair, Telegie transmitters send packets with 50\% of redundancy.

\subsection{Visualization}
Telegie viewers visualize every depth pixel into a quad and use color pixels to map color on the quads. We chose quads because the geometry for the visualization as depth pixels is arranged in 2D grids. By default, our system operates with a color resolution of 1280x720 and a depth resolution of 640x360. In the following study, we used the half resolution of them, 720x360 and 320x180, to avoid networking issues such as occasional frame drops to complicate our analyses.

\begin{figure}[H]
  \includegraphics[width=\textwidth]{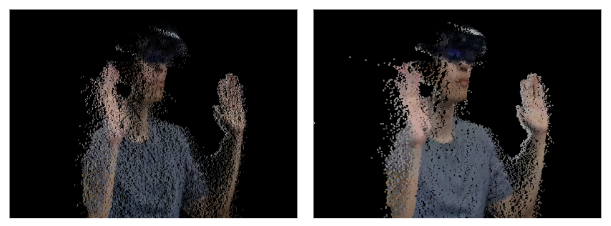}
  \caption{Comparison between a side view of quads facing the center of the camera which originally captured them (left) and another side view with quads rotated towards the user seeing the quads (right).}
  \label{fig:rotation}
\end{figure}

\begin{figure}[H]
  \includegraphics[width=\textwidth]{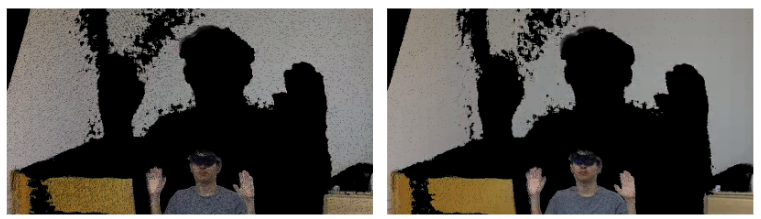}
  \caption{Comparison between without (left) and with (right) quad enlargement by the factor of 1.2.}
  \label{fig:enlargement}
\end{figure}

The quads corresponding to depth pixels get rotated toward the users to enhance their visibility. Without additional rotations, there can be wide gaps between the quads. While these gaps can be seen as natural, reducing these gaps improves visibility especially when the user is seeing the quads from the side (see Figure~\ref{fig:rotation}). Quads are also scaled by the factor of 1.2 to further increase their visibility (See Figure~\ref{fig:enlargement}).

\section{Study}
The present study examined the effectiveness of VC and TP in the context of remote physical therapy. There were two independent variables with two conditions each: media type (VC vs. TP) and the number of cameras used to capture and render the scene (one camera vs. two cameras). Under the VC and TP conditions, physical therapists and patients communicated through the use of Google Meet and Telegie respectively. Depending on the camera condition, views of the patient were captured by either one or two cameras. We compared each condition to investigate their potential as a tool for remote physical therapy.

\subsection{Hypotheses and Research Question}
\label{sec:study_hypotheses}

Based on previous literature~\cite{bystrom1999collaborative,kim2013can, markowitz2018immersive}, we predicted that TP and introducing a second camera would positively affect four outcomes: interpersonal communication responses from therapists, interpersonal communication responses from patients, therapist assessments, and physical therapy evaluations from patients. Our hypotheses were:

\begin{itemize}
  \item H1: The level of interpersonal communication responses from therapists will be higher in TP than in VC.
  \item H2: The level of interpersonal communication responses from patients will be higher in TP than in VC.
  \item H3: The level of therapist assessments will be higher in TP than in VC.
  \item H4: The level of physical therapy evaluations from patients will be higher in TP than in VC.
  \item H5: The level of interpersonal communication responses from therapists will be higher with an additional camera.
  \item H6: The level of interpersonal communication responses from patients will be higher with an additional camera.
  \item H7: The level of therapist assessments will be higher with an additional camera.
  \item H8: The level of physical therapy evaluations from patients will be higher with an additional camera.
\end{itemize}

Additionally, we explored how differences between participants might have influenced the relationship between their study condition and the four outcomes listed above. In particular, we expanded our analysis to account for therapist identity, prior VR experience, prior physical therapy experience, spatial ability of the patients, perceived video clarity, and perceived patient motivation. In doing so, we formed the following research question:

\begin{itemize}
  \item RQ1: Do individual differences between participants affect the relationship between the communication system and the four experimental outcomes?
\end{itemize}

\subsection{Method}
\subsubsection{Participants}
There were two types of participants: therapists and patients. Therapists consisted of 11 physical therapy students enrolled in a clinical Doctor of Physical Therapy program at a US East Coast University. Each therapist completed coursework relevant to their role in this study, such as therapeutic exercise. The average age of the therapists was 24.455 years old (SD = 1.508). Eight of them were female and three of them were male. 

Seventy-six participants from a US West Coast University took on the role of patients. The average age was 25.711 years old (SD = 6.685) across 47 female and 29 male patients. Nineteen patients were assigned to the one-camera VC condition (VC1), 20 patients to the one-camera TP condition (TP1), 19 patients to the two-camera VC condition (VC2), and 18 patients to the two-camera TP condition (TP2). The recruitment and experiment processes were approved by each university’s IRB. 

\subsubsection{Materials and Apparatus}
Patients viewed therapists through an iPad tablet regardless of condition. Depending on the media type condition, therapists viewed patients through a TV monitor or Meta Quest 2 VR headset. In particular, therapists saw patients through a TV monitor in the VC conditions and through a VR headset in the TP conditions. Correspondingly, video streams of patients were captured by webcams in the VC conditions and by Azure Kinect RGBD cameras in the TP conditions. Both patients and therapists used microphones placed in front of them to capture audio input. 

\subsubsection{Design and Procedure}
We used a 2x2 factorial design for our study. The independent variables were media type (VC vs. TP) and the number of cameras (one camera vs. two cameras). Due to the different sample sizes of therapists and patients, we conducted a within-participant study for therapists and a between-participant study for patients. Each session consisted of one therapist and one patient. Each therapist was assigned to eight sessions, two sessions per each of the four experimental conditions. We ordered the conditions of the therapists’ first four sessions based on a 4x4 Latin square. The conditions of their remaining four sessions were repeated in the reverse order (see Table~\ref{tab:latin}). The condition order for the first four therapists were repeated by the remaining therapists. 

\begin{table}[ht]
  \centering
  \begin{tabular}{ccccc}
    \toprule
    & Therapist 1 & Therapist 2 & Therapist 3 & Therapist 4\\
    \hline
    Session 1 & VC1 & VC2 & TP1 & TP2\\
    Session 2 & VC2 & TP1 & TP2 & VC1\\
    Session 3 & TP2 & VC1 & VC2 & TP1\\
    Session 4 & TP1 & TP2 & VC1 & VC2\\
    Session 5 & TP1 & TP2 & VC1 & VC2\\
    Session 6 & TP2 & VC1 & VC2 & TP1\\
    Session 7 & VC2 & TP1 & TP2 & VC1\\
    Session 8 & VC1 & VC2 & TP1 & TP2\\
    \bottomrule
  \end{tabular}
  \caption{The order of experimental conditions assigned to the sessions. The condition order for the first four therapists were repeated by the remaining therapists.}
  \label{tab:latin}
\end{table}

Before the experiment, all participants reported demographic information, prior VR experience, and prior physical therapy experience on a questionnaire. This questionnaire also included a test that measured their spatial ability. In addition to providing this information, therapists spent 10 minutes familiarizing themselves with the systems associated with each experimental condition. Patients did not go through this step because they always saw their therapist through a VC system on a tablet which required no training.

\begin{figure}[H]
  \centering
  \includegraphics[width=0.5\textwidth]{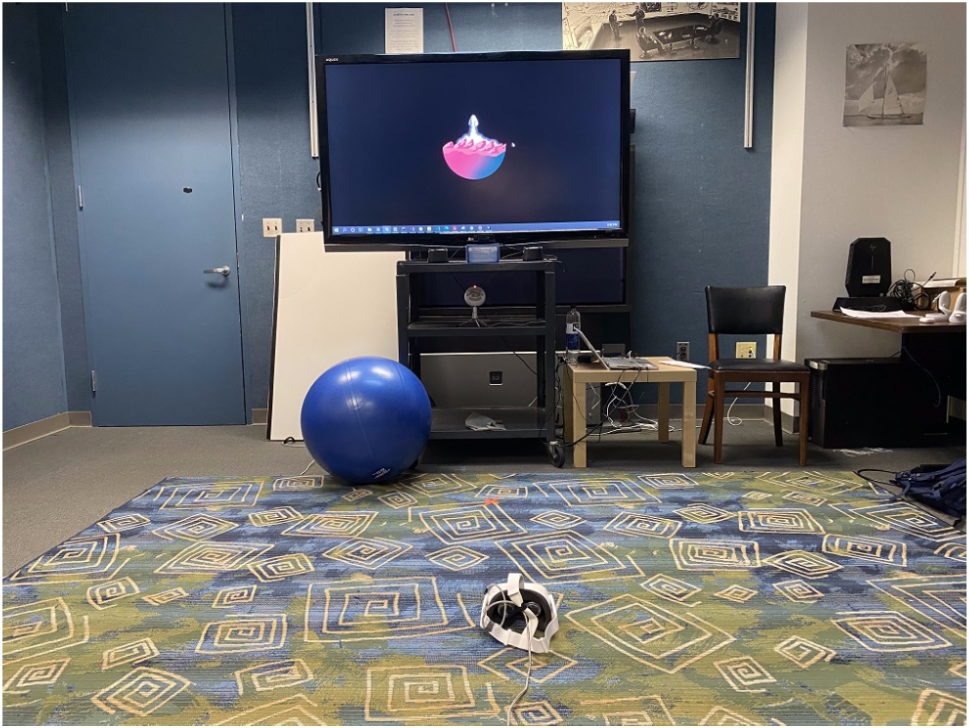}
  \caption{The environment for the therapists had a webcam placed in front of the therapists and a monitor for the therapists to see the patients in VC conditions.}
  \label{fig:unclab}
\end{figure}

\begin{figure}[H]
  \includegraphics[width=\textwidth]{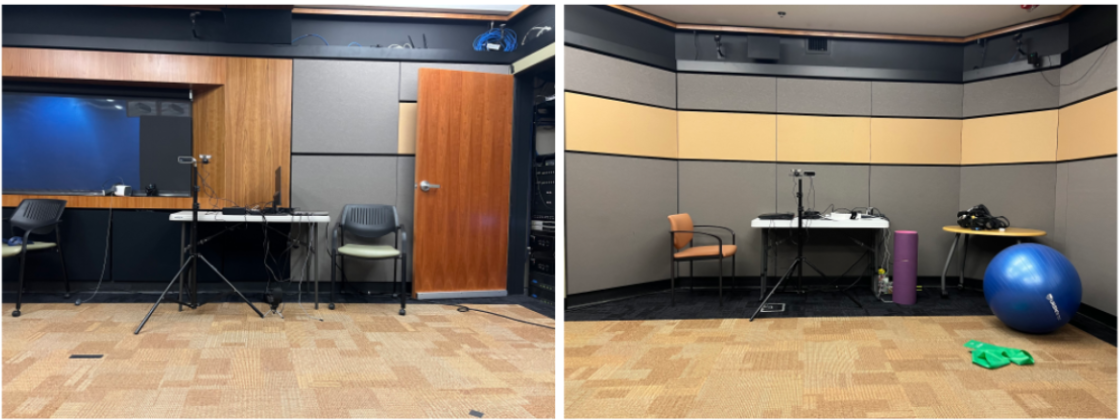}
  \caption{Photos of the front (left) and side (right) cameras of the environment for the patient. There are both a webcam and a RGBD camera installed both from the front and right side of the patients.}
  \label{fig:stanfordlab}
\end{figure}

Our study took place across two labs that were configured to accommodate the experimental conditions. The lab environment for the therapists included one camera that was placed in front them (See Figure~\ref{fig:unclab}). The lab environment for the patients included two web cameras for VC and two RGBD cameras for TP. One of each type of camera was placed in front of the patients, and the remaining two were placed to their right at a 90-degree angle (See Figure~\ref{fig:stanfordlab}.)

\begin{figure}[H]
  \includegraphics[width=\textwidth]{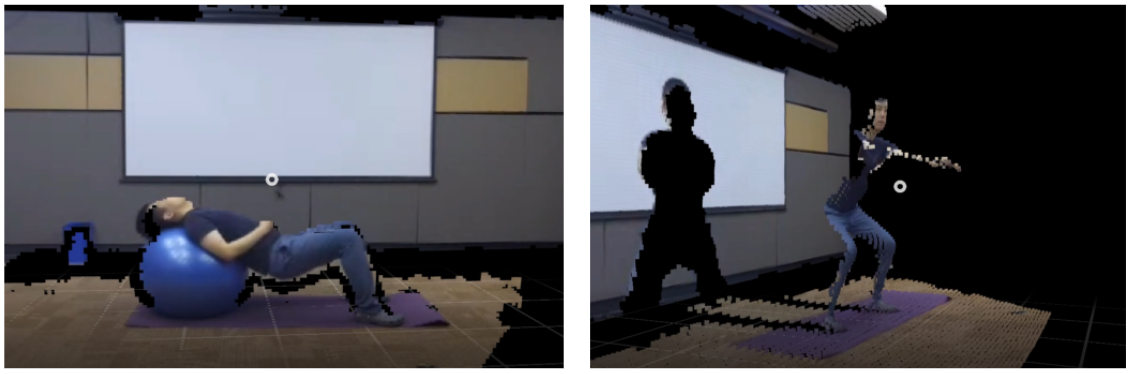}
  \caption{Captures of Telegie from a user's perspective wearing a VR headset in the TP1 condition.}
  \label{fig:tp_captures}
\end{figure}

Visuals for patients and therapists varied based on condition. In the VC1 condition, patients and therapists saw video streams coming from their front-facing web cameras. In the VC2 conditions, patients and therapists saw video streams coming from both their front and side-facing web cameras. This meant that the VC conditions, participants saw both each other and themselves. In the TP conditions, therapists wore a VR headset to view their patient and were thus unable to see themselves. Depending on the camera condition, therapists could see a video stream of their patient constructed by either the front-facing RGBD camera alone (TP1) or both the front and side-facing RGBD cameras (TP2).  Figure~\ref{fig:tp_captures} demonstrates how patients looked to therapists in the TP1 condition.

\begin{figure}[h]
 \centering 
 \includegraphics[width=\columnwidth]{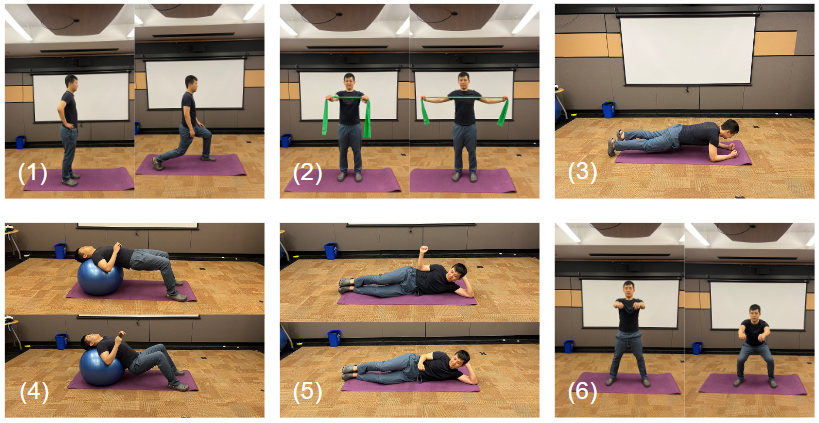}
 \caption{Every session consists of six exercises: (1) lunge, (2) elastic band bilateral horizontal abduction, (3) plank, (4) ball bridge upper back, (5) side lying external rotation, and (6) squat.}
 \label{fig:exercises}
\end{figure}

During each 15-minute session, therapists instructed patients to complete six exercises for two sets (see  Figure~\ref{fig:exercises}.) We selected a diverse set of exercises that could be evaluated spatially. For example, during a plank exercise, a physical therapist might want to check if their patient’s back is straight. After each session, both the therapists and patients completed a post-questionnaire.

The experimenter interviewed each therapist after they completed their eight remote physical therapy sessions. Interview questions prompted therapists to compare their experiences across each experimental condition. 

\subsubsection{Measures}
\paragraph{Prior VR Experience}
Participants indicated whether or not they had prior VR experience. Two out of the 11 therapists and 49 out of the 76 patients indicated that they previously experienced VR.

\paragraph{Prior Physical Therapy Experience}
Patients indicated whether or not they had prior physical therapy experience. Forty-five out of the 76 patients indicated that they previously experienced physical therapy.

\paragraph{Spatial Ability}
Spatial ability was measured through a mental rotation test~\cite{shepard1971mental,peters1995redrawn}. This test consisted of five questions that asked participants if two representations of objects were the same but in a different orientation. The average score for therapists was 4.545 out of 5 (SD = 0.688), and the average score for patients was 4.500 (SD = 0.825). 

\paragraph{Interpersonal Communication}
After their remote physical therapy session, we asked participants to rate the levels of social presence~\cite{herrera2020effect}, communication satisfaction~\cite{oh2019effects}, interpersonal liking~\cite{oh2019effects}, and Inclusion of Other in the Self (IOS)~\cite{aron1992inclusion} they experienced during their session. Due to their high correlation with one another, these four measures were averaged together into one interpersonal communication score (Cronbach's $\alpha$ = 0.758 for therapists; Cronbach's $\alpha$ = 0.788 for patients). When taking the average, IOS was rescaled from a 1-7 scale to a 1-5 scale to match the 5-point Likert scale of the other three measures.

\paragraph{Video Clarity}
Participants rated how clear the video stream was after each session on a 5-point Likert scale. We captured this metric because video quality between the VC system and TP system differed. The VC system was a mature commercial system whereas the TP system was more experimental. The mean value of perceived video clarity was 3.066 (SD = 1.215) across therapists and 3.829 (SD = 0.806) across patients.

\paragraph{Perceived Patient Motivation}
Therapists rated how motivated each of their patients were during their session on a 5-point Likert scale. The mean value of perceived patient motivation was 3.829 (SD = 0.839).

\paragraph{Therapist Assessments}
Therapists rated how accurately and quickly their patients completed each exercise on a 5-point Likert scale. Due to the high correlation between accuracy and quickness (Cronbach's $\alpha$ = 0.917), we combined the two measures into a therapist assessment score. The mean value of therapist assessment was 4.427 (SD = 0.543).

\paragraph{Physical Therapy Questionnaire for Patients}
Patients answered 16 questions from \citeA{bailenson2008effect} about their physical experience after completing their session. These questions asked patients to rate their impressions of their therapist, environment, and tasks on a 5-point Likert scale. The 16 questions were highly correlated to each other (Cronbach's $\alpha$ = 0.895) and were averaged into a physical therapy evaluation score. Higher scores were associated with more positive valence. The mean value of physical therapy evaluation was 3.623 (SD = 0.545).

\paragraph{Interviews}
Interviews with each therapist were conducted at the end of their eight sessions by an experimenter at the site for therapists. Each interview lasted around 10 minutes. Therapists were asked a series of questions and follow-up questions about comparing their experiences with each experimental condition. Some examples of questions include ``compare TP sessions to VC sessions.'' Interviews were audio-recorded, transcribed using transcription software, and thematically analyzed.

\section{Results}
In this section, we report the findings from our experiment. First, we describe the relationships between each experimental condition and the four measures of theoretical interest. Next, we expand this analysis to account for differences between participants. Finally, we summarize insights gained from therapist interviews and open-ended responses from patients.

\begin{figure}[H]
  \begin{subfigure}{0.49\columnwidth}
    \includegraphics[width=\columnwidth]{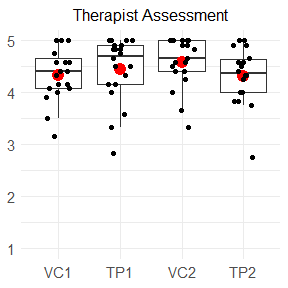}
  \end{subfigure}
  \begin{subfigure}{0.49\columnwidth}
    \includegraphics[width=\columnwidth]{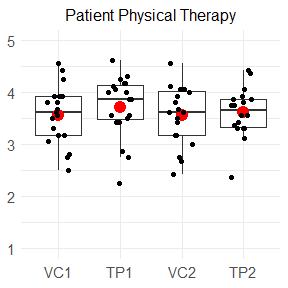}
  \end{subfigure}\\
  \begin{subfigure}{0.49\columnwidth}
    \includegraphics[width=\columnwidth]{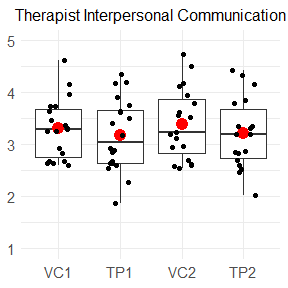}
  \end{subfigure}
  \begin{subfigure}{0.49\columnwidth}
    \includegraphics[width=\columnwidth]{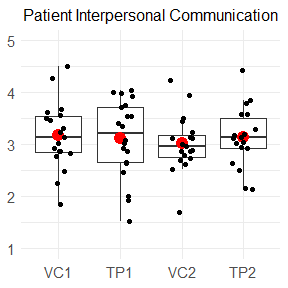}
  \end{subfigure}
  \caption{The distributions of the four dependent variables across the experimental conditions.}
  \label{fig:extension_box}
\end{figure}

Figure~\ref{fig:extension_box} provides the distributions of all four dependent variables across experimental conditions. Table~\ref{tab:therapist_pt_mean} shows the mean and standard deviation of therapist assessments of how each exercise was performed by patients. The accuracy and quickness of their performance were highly correlated with each other in the positive direction ($\rho$ = 0.788).

We tested all eight hypotheses proposed in Section~\ref{sec:study_hypotheses}, but none of them were statistically significant. Therefore, there were no significant differences between using TP over VC and using two cameras over one camera during the remote physical therapy sessions. Table~\ref{tab:postsurvey_hypotheses} summarizes these results. 

\begin{table}[ht]
  \centering
  \begin{tabular}{ccc}
    \toprule
    Exercises & Accuracy & Quickness\\
    \hline
    Lunge & 4.408 (0.751) & 4.368 (0.763)\\
    Band & 4.500 (0.663) & 4.408 (0.715)\\
    Plank & 4.566 (0.660) & 4.553 (0.737)\\
    Ball & 4.197 (0.966) & 4.171 (0.929)\\
    Rotation & 4.487 (0.622) & 4.526 (0.599)\\
    Squat & 4.434 (0.736) & 4.500 (0.792)\\
    \hline
    Average & 4.432 (0.551) & 4.421 (0.572)\\
    \bottomrule
  \end{tabular}
  \caption{The mean values (and standard deviations) of therapist assessments.}
  \label{tab:therapist_pt_mean}
\end{table}

\begin{table}[ht]
  \centering
  \begin{tabular}{ccc}
    \toprule
    & TP & Cameras\\
    \hline
    \makecell{Therapist Interpersonal\\Communication} & \makecell{t(73.451)=-1.025\\p=0.309} & \makecell{t(72.947)=0.408\\p=0.684}\\
    \hline
    \makecell{Patient Interpersonal\\Communication} & \makecell{t(73.061)=0.245\\p=0.807} & \makecell{t(72.238)=-0.449\\p=0.655}\\
    \hline
    \makecell{Therapist Assessment} & \makecell{t(72.057)=-0.578\\p=0.565} & \makecell{t(73.981)=0.407\\p=0.685}\\
    \hline
    \makecell{Patient Physical\\Therapy Evaluations} & \makecell{t(73.836)=0.854\\p=0.396} & \makecell{t(73.870)=-0.434\\p=0.665}\\
    \bottomrule
  \end{tabular}
  \caption{The t-values and p-values of statistical tests corresponding to the eight hypotheses.}
  \label{tab:postsurvey_hypotheses}
\end{table}

To understand why our hypotheses were not supported, we conducted an exploratory analysis and used linear mixed models to control for individual differences between participants that might have influenced our experimental outcomes. In particular, we explored the effects of therapist identity, video clarity, spatial ability, perceived patient motivation, gender, prior VR experience, and prior PT experience. We carried out this analysis using lme4 1.1-27.1~\cite{bates2014fitting}. In the following section, we control for these additional variables and assess their effect on therapist assessment scores. We then repeat this analysis for the remaining three dependent variables: physical therapy evaluations from the patients, interpersonal communication responses from the therapists, and interpersonal
communication responses from the patient.

\subsection{Therapist Assessments}
\label{sec:results_rq_evaluations}

\begin{figure}[ht]
\includegraphics[width=\columnwidth]{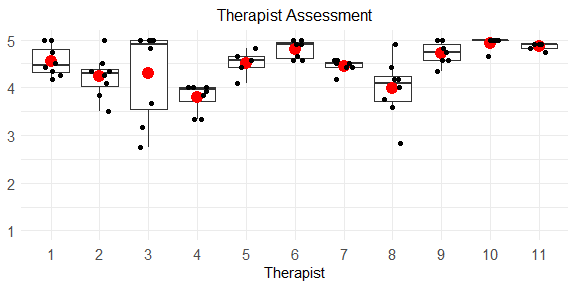}
  \caption{The distributions of assessment scores from each of the therapists.}
  \label{fig:therapist_evalution_therapists}
\end{figure}

Figure~\ref{fig:therapist_evalution_therapists} visualizes the cross-therapist variation in assessment scores. Individual differences between both therapists and patients might have influenced therapist assessments. This was examined using a linear mixed model with the experimental conditions as a fixed effect and the therapist identity as a random effect. Based on this model, using two cameras had a marginally significant positive effect  (b = 0.246, p = 0.093) on assessment scores and the interaction between TP and using two cameras had a marginally significant negative effect (b = -0.358, p = 0.085) on assessment score.

\subsubsection{Video Clarity}
\label{sec:results_rq_video_clarity}

\begin{figure}[ht]
  \begin{subfigure}{0.49\columnwidth}
    \includegraphics[width=\columnwidth]{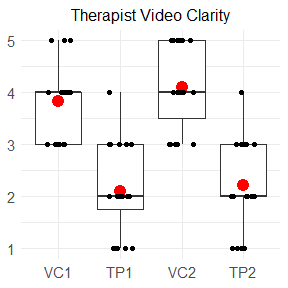}
  \end{subfigure}
  \begin{subfigure}{0.49\columnwidth}
    \includegraphics[width=\columnwidth]{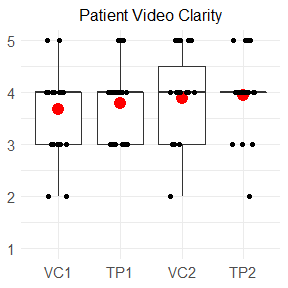}
  \end{subfigure}
  \caption{The distributions of video clarity levels reported by therapists and patients per experimental condition.}
  \label{fig:video_clarity_boxes}
\end{figure}

\begin{figure}[ht]
  \begin{subfigure}{0.49\columnwidth}
    \includegraphics[width=\columnwidth]{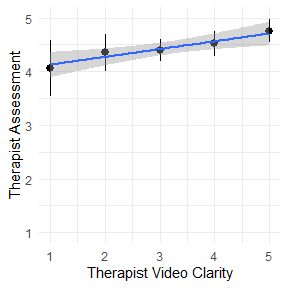}
  \end{subfigure}
  \begin{subfigure}{0.49\columnwidth}
    \includegraphics[width=\columnwidth]{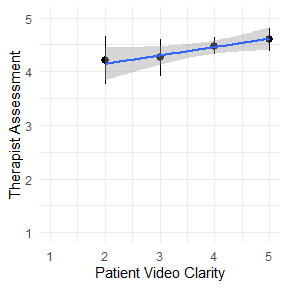}
  \end{subfigure}
  \caption{Visualization of the linear models from video clarity levels reported by therapists and patients on therapist assessments.}
  \label{fig:video_clarity_slopes}
\end{figure}

Due to the greater technical maturity of the VC system (Google Meet) compared to the TP system (Telegie), the TP system had a noticeably lower video resolution. This was reflected by questionnaire responses, where clarity levels reported by the therapists were highly correlated in the negative direction ($\rho$ = -0.75) with TP. Comparatively, patients, who only interfaced with VC via a tablet, reported video clarity levels that did not show any correlation ($\rho$ = 0.05). Figure~\ref{fig:video_clarity_boxes} shows the distribution of video clarity per conditions. Figure~\ref{fig:video_clarity_slopes} shows the positive slopes on assessment scores from both video clarity levels.

Because differences in video clarity could be attributed to the TP system’s lack of technical maturity, it was worth controlling for perceived clarity levels when examining the relationship between media type and therapist assessments. After adding video clarity as a fixed effect to a linear mixed model representing this relationship, several significant findings were revealed. In particular, TP had a significant positive effect (b = 0.384, p = 0.048) and the interaction between TP and using two cameras had a marginally significant negative effect (b = -0.340, p = 0.097) on therapist assessments. Using two cameras did not have a significant effect (b = 0.177, p = 0.226). Our analysis also revealed that video clarity levels reported by the therapists had a significant positive effect (b = 0.169, p = 0.026) on assessment scores and video clarity level reported by the patients had a marginally significant positive effect (b = 0.119, p = 0.078) on assessment scores.

\subsubsection{Spatial Ability}
\label{sec:results_rq_spatial_ability}

\begin{figure}[ht]
  \begin{subfigure}{0.49\columnwidth}
    \includegraphics[width=\columnwidth]{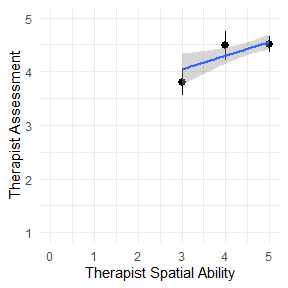}
  \end{subfigure}
  \begin{subfigure}{0.49\columnwidth}
    \includegraphics[width=\columnwidth]{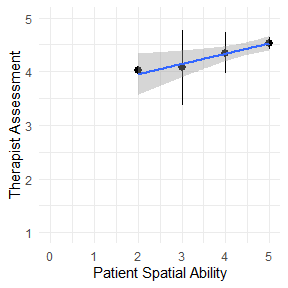}
  \end{subfigure}
  \caption{Visualization of the linear models from spatial ability levels of therapists and patients to therapist assessments.}
  \label{fig:spatial_ability_slopes}
\end{figure}

Figure~\ref{fig:spatial_ability_slopes} shows the linear relationship between spatial ability and therapist assessment for both therapists and patients. Whereas most physical therapists achieved high spatial ability scores, the greater variance in spatial ability amongst patients might have impacted their therapy assessment scores. After controlling for patient spatial ability, our linear mixed model no longer produced a significant relationship between experimental condition and therapist assessment. However, there was a significant positive effect (b = 0.155, p = 0.021) of spatial ability on assessment. 

\subsubsection{Perceived Patient Motivation}
\label{sec:results_rq_patient_motivation}

\begin{figure}[ht]
  \begin{subfigure}{0.49\columnwidth}
    \includegraphics[width=\columnwidth]{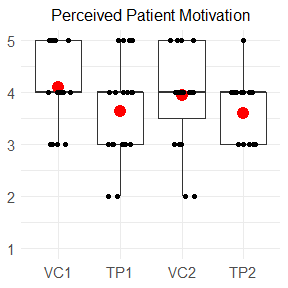}
  \end{subfigure}
  \begin{subfigure}{0.49\columnwidth}
    \includegraphics[width=\columnwidth]{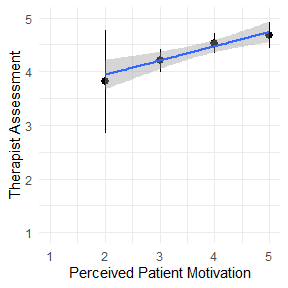}
  \end{subfigure}
  \caption{The distributions of perceived patient motivation levels reported by therapists per condition (left) and the linear model from perceived patient motivation levels reported by therapists to therapist assessments (right).}
  \label{fig:patient_motivation_box}
\end{figure}

The effect of perceived patient motivation reported by therapists on assessment outcomes was also explored. Perceived patient motivation was negatively correlated with TP ($\rho$ = -0.24) and positively correlated with therapist assessments ($\rho$ = 0.41). Figure~\ref{fig:patient_motivation_box} shows the distributions of reported motivation levels by experimental condition and the relationship between motivation levels and assessment scores.

To examine the effects of the experimental conditions when controlling for perceived motivation, it was added as a fixed effect to a linear mixed model. This model showed that using two cameras (b = 0.275, p = 0.062) and the interaction between TP and using two cameras (b = -0.391, p = 0.060) had effects on therapist assessment. However, these effects were in the same direction as the version of the model without the motivation level controlled. Moreover, perceived motivation level had a significant positive effect on therapist assessment. 

The effects of the gender, prior VR experience, and prior PT experience were also investigated by adding them as additional fixed effects. However, no statistically significant effect was found. 

\subsection{Expansion to Other Dependent Variables}
\label{sec:results_rq_expansion}

\begin{table}[h]
  \centering
  \scalebox{0.7}{
    \begin{tabular}{cccccc}
      \toprule
      \makecell{Additional\\Fixed Effect\\(Corresponding\\Section)}& \makecell{Fixed Effect}& \makecell{Therapist\\Assessment} & \makecell{Patient\\Physical\\Therapy} & \makecell{Therapist\\Interpersonal\\Communication} & \makecell{Patient\\Interpersonal\\Communication}\\
      \hline
      \multirow{3}{*}{\makecell{None\\(\ref{sec:results_rq_evaluations})}} & TP & 0.101 & 0.162 & -0.130 & -0.034\\
      & Cameras & 0.246$^\dagger$ & 0.002 & 0.088 & -0.163\\
      & TP $\times$ Cameras & -0.357$^\dagger$ & -0.093 & -0.006 & 0.181\\
      \midrule
      \multirow{4}{*}{\makecell{Therapist\\Video Clarity\\(\ref{sec:results_rq_video_clarity})}} & TP & 0.403* & 0.329 & 0.384* & 0.043\\
      & Cameras & 0.198 & -0.026 & 0.008 & -0.175\\
      & Therapist Video Clarity & 0.173* & 0.098 & 0.297* & 0.045\\
      & TP $\times$ Cameras & -0.341 & -0.080 & 0.024 & 0.186\\
      \midrule
      \multirow{4}{*}{\makecell{Patient\\Spatial Ability\\(\ref{sec:results_rq_spatial_ability})}} & TP & 0.039 & 0.174 & -0.157 & 0.004\\
      & Cameras & 0.191 & 0.012 & 0.064 & -0.137\\
      & Patient Spatial Ability & 0.155* & -0.030 & 0.069 & -0.074\\
      & TP $\times$ Cameras & -0.266 & -0.110 & 0.035 & 0.138\\
      \midrule
      \multirow{4}{*}{\makecell{Perceived\\Patient Motivation\\(\ref{sec:results_rq_patient_motivation})}} & TP & 0.202 & 0.198 & 0.088 & -0.056\\
      & Cameras & 0.275$^\dagger$ & 0.012 & 0.154 & -0.172\\
      & Perceived Patient Motivation & 0.225* & 0.089 & 0.496* & -0.054\\
      & TP $\times$ Cameras & -0.391$^\dagger$ & -0.103 & -0.080 & 0.189\\
      \bottomrule
    \end{tabular}
  }
  \caption{Estimated slopes and their p-values of the fixed effects from the linear mixed models as the expansion of Section~\ref{sec:results_rq_evaluations} to all four dependent variables. Each column represents a dependent variable. (*: p $<$ 0.050, $^\dagger$: p $<$ 0.100)}
  \label{tab:extension_models}
\end{table}

In this section, we expand our analysis to explore how differences between participants might have shaped physical therapy evaluations from the patients, interpersonal communication responses from the therapists, and interpersonal communication responses from patients. Similar to the previous section, we use linear mixed models with fixed effects to accomplish this. Table~\ref{tab:extension_models} summarizes results from this analysis. Therapist video clarity levels and perceived participant motivation had an effect on interpersonal communication responses from therapists. However, patient spatial ability did not influence interpersonal communication responses. No significant effect was found for both patient responses to physical therapy and interpersonal communication as dependent variables.

\begin{figure}
  \includegraphics[width=\textwidth]{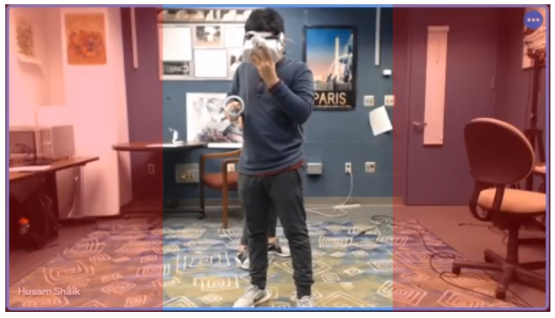}
  \caption{Visualization of regions the independent rater counted the therapists as watching the patients from the side camera. Rater was asked to consider the therapists as watching the patients from sides when the therapists entered the red area.}
  \label{fig:tp2_side_region}
\end{figure}

Unlike VC, using two cameras with the TP system did not result in positive outcomes compared to using one camera for TP. To better understand this finding, an independent rater watched video recordings of the TP sessions with two cameras. They reported the ratio of time that therapists took advantage of the second camera by observing their patients from the side. Figure~\ref{fig:tp2_side_region} visualizes the regions of the therapist’s environment, according to the rater, that therapists were able to view angles of patients captured by the side camera. Their ratings indicated that therapists watched patients from the side for only 13.546\% of the time during the two-camera TP sessions. During seven sessions out of the 18 two-camera TP sessions, therapists used the side camera less than 10\% of the time. 

\subsection{Open-ended Responses}
\label{sec:results_interviews}

\subsubsection{Therapist Interviews}

All therapists experienced all four experimental conditions during their participation in the present study. After spending three hours conducting remote physical therapy sessions across each condition, they described their experiences with each system in an interview. 

The most common issue of TP mentioned by all therapists was that they could not see themselves. Therapist \#10 said, "[I]f I was asked to do like a completely new exercise, it would be very hard." Therapist \#11 said, "I will say that, surprisingly, [it was] easier to show the exercises without the VR because I could also see myself in the zoom."

Another TP issue mentioned by 10 therapists was insufficient resolution of depth pixels. This issue occurred because in order to guarantee low network latency for the entire duration of the study, resolution was downsampled into half for both width and height. Therapist \#3 said, "around the knees, I can really only see like a knee cap."

Two therapists compared the issue of not being able to see themselves and to the issues with pixelation. Both therapists found the insufficient resolution to be the greater problem. Therapist \#8 said, "I think the pixelation was more frustrating because I just couldn't really fully see them and all their different positions."

Four therapists said VC was better than TP. For example, Therapist \#4 said "And I will say that with a VR, it definitely was harder. Like with it being pixelated, it was harder from my mindset of a therapist to pick out the things that I want to fix because it just wasn't quite as clear… I can't tell if I need to have them move their feet a little bit further apart, or if that like positioning wise, or if that what I'm seeing is what they're actually doing."

Five therapists said the two-camera TP system was better than the one-camera TP system and three mentioned that they felt similar. Describing TP2 as more preferred, Therapist \#6 said, "I did notice the two cameras system. One, even though we're still having them turn to their side it was a clear image."

Seven therapists mentioned that they preferred VC2 over VC1, and two therapists said they preferred VC2 the most out of all conditions. Therapist \#3, while mentioning their preference for VC2 over VC1, said, "When I was first doing the two cameras, I was almost exclusively looking at head on camera. So I feel like I was underutilizing the second camera. Especially with the static movements. Once we were getting into like the squat and lunge and stuff like that. It was kind of nice to have that second camera to see the side, which can be adjusted. You can just make the patient rotate… But you can't see real time. So I think it was helpful to have the second camera but maybe not like essential." 

Two therapists said they did not find VC2 much better than VC1. Therapist \#7 said, "Honestly, I was fine with the one video like the one camera. I don't think the second camera added all that much. Like there was a couple times where, you know, it was helpful to have like ask the participant to go on a couple different positions. But I don't feel like there was that it only took like a couple seconds. And it wasn't didn't seem like too much of a hassle."

Although six therapists said they experienced greater social presence in TP. Therapist \#5 said, "[Y]ou almost want to like reach towards the patient." However, Therapist \#9 felt that wearing a headset made them feel less socially present. While describing their reasoning, they said, "because of just of the pixels and the fact that I didn't like that half of my face was covered and goggles when I was trying to talk to the patient."

On the topic of how TP could be improved, five therapists mentioned that having higher resolution would enhance the experience. Therapist \#9 said, "I don't know if that's a cost and benefit type thing. But I did think about that. If it could be more high resolution that would be that would be even better." 

Five therapists stated that having more VR experience could be helpful for using TP systems. Therapist \#3 said, "I think it was a lot more effective towards the end from having practice. But also, yeah, I think I just felt more comfortable with the VR in general."

One therapist said the assessment questions did not work well since many patients already knew the exercises. The therapist said, "I think just a, maybe just a broad comment about the survey. It did seem like a lot of the participants came in with a pretty good understanding of most exercises. And so the question asking how well they learned the exercise didn't seem to be very representative of the situation because if they already came in knowing the exercise, there wasn't a great way to answer how well, or quickly, they learned it."

\subsubsection{Patient Open-ended Responses}
Patients had the opportunity to optionally share additional comments at the end of their post-questionnaire. Since each patient only experienced one condition, their comments did not include the comparison between conditions.

Nineteen out of the 76 patients pointed out that the tablet they used for seeing the therapist was too small or was placed too low. Two out of the 76 patients mentioned that the audio quality between them and their therapists was poor.

\section{Discussion}
Based on previous literature~\cite{bystrom1999collaborative,kim2013can, markowitz2018immersive}, we expected that using a TP system and two-camera system for remote physical therapy would have positive effects on four dependent variables: interpersonal communication responses from therapists, interpersonal communication responses from patients, therapist assessments, and physical therapy evaluations. However, none of our eight hypotheses that predicted these positive outcomes were statistically significant. To better understand our findings, we controlled for individual differences that might have influenced the relationship between the systems and outcomes of interest. 

We first examined the effect of these differences on therapist assessment in Section~\ref{sec:results_rq_evaluations}. After controlling for therapist identity, we found a positive marginally significant effect of using two cameras and a negative marginally significant effect from the interaction between TP and using two cameras on therapist assessments. Controlling for video clarity resulted in a significant positive effect of TP on assessment scores. A positive significant effect of video clarity, patient spatial ability, and perceived patient motivation on therapist assessment was also found. 

After expanding this analysis to the remaining three dependent variables in Section~\ref{sec:results_rq_expansion}, our findings were partially replicated for interpersonal communication responses from therapists. TP had a significant positive effect on interpersonal communication responses when controlling for video clarity. However, controlling for spatial ability did not produce a significant effect. 

In therapist interviews, most of them found the low resolution (10 out of 11 therapists) and the inability to see themselves (all 11 therapists) in TP systems to be an issue. Four therapists stated they preferred TP over VC. Five therapists said they preferred using two cameras for TP over having one camera and seven therapists said they preferred having the additional camera for VC. This preference to use a VC system with two cameras over one camera was also shown in results from the linear mixed model in Section~\ref{sec:results_rq_evaluations} regarding the experimental conditions on therapist assessments. Finally, interviews revealed that many therapists preferred VC over TP.

Our study leveraged the expertise of physical therapists to evaluate the potential of TP systems for remote physical therapy. Insights gained from these experts allowed us to pinpoint some of the affordances and challenges of incorporating TP within this realistic context.

\section{Conclusion}
In this paper, we examined the effect of using a TP system vs a VC system and using two cameras vs one camera for remote physical therapy. Our study took place synchronously across two locations where participants who were actual physical therapists communicated remotely with participants who were assigned as patients. Given this context and our goal to rigorously examine the different remote communication systems, a large number of participants were recruited. 

Our key findings are as follows: With the individual differences between both therapists and patients controlled, the condition using the VC system with two cameras was found to be better than other experimental conditions. When we controlled for reported levels of video clarity, using the TP system was found to be better than using the VC system. Patients' spatial ability level performed very well as a predictor of physical therapy session evaluations from the therapists. And finally, perceived motivation levels of the patients reported by therapists significantly explained physical therapy evaluations from therapists.

\subsection{Limitations}
One limitation was that the video resolution of Telegie, the TP system, was relatively poor. To provide stable streaming, we downsampled both the color and depth pixels captured by the Azure Kinect device into half per dimension.

Another limitation was that assessment scores from the therapists likely suffered from a ceiling effect. Average scores were above a 4 out of 5. Furthermore, a more difficult set of exercises should have been chosen.

The sample size of the study could have also been larger. We were unable to achieve our initial recruitment goal of 88 participants due to the difficulty of setting up a remote dyadic study across different time zones with one in the dyad needing to be a physical therapist. With a larger sample size, the marginally significant effects in Section~\ref{sec:results_rq_evaluations} could have been analyzed more clearly.

\subsection{Future Directions}
In this study, due to the difficulty of setting up the TP system, we did not find a bidirectional TP condition to be realistic. As a result, only the therapists wore VR headsets. The setup of the TP system can also be simplified. For example, calibration of the system when working with multiple cameras should be automatic. Future research should investigate bidirectional and simplified TP systems. 

While we found patients' spatial ability to predict their task performance evaluated by therapists, there might be additional measures of patient ability worth capture. 

Measuring the impact of providing sufficient TP experience to the therapists can also be seen as a direction to explore. In our study, most of the therapists were not only new to the TP system, but also did not have prior experience using VR. With more VR experience, therapists might better utilized information from the side cameras in TP.

\subsection{Implications}
From this study, video clarity levels reported from the participants produced the most salient findings. After we controlled for video clarity levels across the experimental variables, TP had a positive effect on physical therapy assessment scores and interpersonal communication. It is also worth noting that 10 out of the 11 therapists mentioned that low resolution was an issue of TP in their interviews. Based on these observations, it is clear that fidelity is a factor that TP systems should not take lightly. When building systems for remote physical therapy, the goal should be to have enough resolution to capture facial expressions and the exact positions of the limbs. 

Another lesson that was revealed from interviews is the importance of letting the TP users see themselves. This suggests that using augmented reality technology as an alternative to VR headsets may improve the user experience of TP systems.

\pagebreak
\bibliographystyle{apacite}
\bibliography{main.bib}

\end{document}